\documentclass[12pt]{iopart}
\usepackage{natbib}
\usepackage{epsfig}

\newcommand{\av}[1]{\langle {#1}\rangle}
\newcommand{\fluck}[1]{\langle {#1}^2\rangle}

\newcommand{\be}{\begin{equation}}
\newcommand{\ee}{\end{equation}}

\newcommand{\binom}[2]{\left(\begin{array}{*{20}c} {#1} \\ {#2} \\ \end{array}\right)}
\begin{document}

\title{Zero temperature Glauber dynamics on complex networks}

\author{Claudio Castellano$^1$ and  Romualdo Pastor-Satorras$^2$}
\address{$^1$ CNR-INFM, SMC, Dipartimento di Fisica, Universit\`a di Roma
``La Sapienza'', P. le Aldo Moro 2, I-00185 Roma, Italy and
CNR, Istituto dei Sistemi Complessi, Roma, Italy}
\address{$^2$ Departament de F\'\i sica i Enginyeria Nuclear, Universitat
  Polit\`ecnica de Catalunya, Campus Nord B4, 08034 Barcelona, Spain}

\begin{abstract}
  We study the Glauber dynamics at zero temperature of spins placed on
  the vertices of an uncorrelated network with a power-law degree
  distribution.  Application of mean-field theory yields as main
  prediction that for symmetric disordered initial conditions the mean
  time to reach full order is finite or diverges as a logarithm of the
  system size $N$, depending on the exponent of the degree
  distribution.  Extensive numerical simulations contradict these
  results and clearly show that the mean-field assumption is not
  appropriate to describe this problem.
\end{abstract}


\maketitle

\section{Introduction}

The zero temperature Glauber dynamics for the Ising model is one of
the simplest ordering dynamics that can be devised for a set of
spin-like variables.  Its sole physical ingredient is the tendency of
each spin to align with the majority of its neighbors, and for this
reason Glauber dynamics has been long investigated as a model for the
ordering of magnets as well as of other systems where the evolution is
driven by the minimization of interfacial energy without conservation
constraints~\cite{Bray94}.  Despite its simplicity, the zero
temperature Glauber dynamics exhibits nontrivial phenomena already on
regular lattices.  In $d=1$ its ordering process can be mapped to an
ensemble of coalescing random walkers and, as a consequence, the mean
ordering time grows with the system size $N$ as $N^2$~\cite{Bray90}.
The same diffusive behavior occurs at $T>0$ in higher dimensions, but
for strictly vanishing temperature there is a probability that the
dynamics remains trapped in states not fully
ordered~\cite{Spirin01,Spirin02}.  In $d=2$ the lack of global
ordering occurs about 3 times out of 10, in the limit of large
systems, and it is associated to the formation of two or more ordered
stripes spanning the system along one direction.  In three (and
presumably higher) dimensions the probability of reaching the fully
ordered state goes to zero as the system size increases: domains of
opposite magnetization coexist, with some ``blinkers'' active at the
boundaries~\cite{Spirin01,Spirin02}.

The simplicity of its local interactions makes Glauber dynamics a
natural model for the evolution of some social systems, whose dynamics
can be defined in terms of agents characterized by a binary variable
(for instance their opinion with respect to some issue) that is
updated in response to the pressure of their peers.  While for
physical applications it is appropriate to study spins on regular
$d$-dimensional lattices, the modelization of social systems
immediately raises the question of the behavior of Glauber dynamics
when the interaction pattern is disordered, i.e., agents live on the
vertices of a network \cite{barabasi02}.  In this scenario, the
nontrivial topological attributes of networks, such as the small-world
property \cite{watts98} and the scale-free property \cite{barab99},
are expected to play some relevant role.  Steps in the direction of
investigating these effects have been recently taken for other types
of ordering dynamics, including the voter
model~\cite{Castellano03,Vilone04,Suchecki05,Sood05} and the Axelrod
model~\cite{Klemm03}.

Among the first investigations of Glauber dynamics on networks we can
mention the work of Boyer and Miramontes~\cite{Boyer03}, who found
that the long-range connections of a Watts-Strogatz network
\cite{watts98} induce a pinning of the ordering process when domains
reach a characteristic size, depending on the density of such
connections.  The Watts-Strogatz network is a rather special type of
network, where, despite the small-world effect, the concept of a local
$d$-dimensional regular neighborhood of a site maintains a
significance, at least for a small density of long-range connections.
Other kinds of networks, such as the random (Erd\"os-R\'enyi) network
\cite{erdos59} are instead not a perturbation of a regular lattice,
and also in this case it has been shown that the zero temperature
Glauber dynamics does not lead to full
ordering~\cite{Svenson01,Castellano05}.  However, for networks of
finite size $N$, it occurs that some realizations of the process
actually lead to complete order, while others get trapped in an
infinitely long-lived (meta)stable state. The probability that a run
ends up in a disordered state turns out to grow with $N$: order is not
reached for asymptotically large $N$ \cite{Castellano05}.

Very recently, Zhou and Lipowsky~\cite{Zhou05} have presented a
mean-field (MF) treatment of the zero temperature Glauber dynamics on
generic uncorrelated complex networks whose degree distribution
$P(k)$, defined as the probability that a vertex is connected to other
$k$ vertices, decreases as a power-law, $P(k) \sim k^{-\gamma}$, with a
characteristic degree exponent $\gamma$.  Within this approach, they find
that a dynamic transition takes place for the particular value
$\gamma_c=5/2$.  For $\gamma>5/2$, the system reaches the ordered state in a
time that diverges logarithmically with the system size. For $\gamma<5/2$,
on the other hand, the ordering time is finite and independent from
$N$.  These results are claimed to be confirmed by numerical
simulations.  The results reported in Ref.~\cite{Zhou05} are
remarkable in several respects. In the first place, they seem to
contradict the difficulty observed to reach complete order generally
observed for Glauber dynamics on disordered interaction patterns.
Secondly, the transition in the dynamical properties occurs for a
value of the exponent $\gamma$ that has not been associated to special
behavior for other models on networks with a power-law degree
distribution, in which transitions typically occur for $\gamma=3$
\cite{pv01a,newman00,havlin00}, below which the second moment of the
degree distribution diverges and the network becomes scale-free
\cite{barab99}.

In this paper we present a refined study of the MF theory for the zero
temperature Glauber dynamics presented in Ref.~\cite{Zhou05},
recasting it more conveniently in a rate equation in the continuous
time limit. In our analysis we take into account the finite size
effects of the network in a more precise way, elucidating what happens
at $\gamma=5/2$. In particular, we present a discussion of the differences
shown by the dynamics depending on the initial conditions, that is,
depending on whether the initial state is completely random and
uncorrelated with zero average magnetization (symmetric initial
state), or it corresponds to a state with a given initial, small,
magnetization (asymmetric initial state). In the former case, we
show that a proper description of the ordering process needs to take
into account the initial diffusive contribution of the dynamics, not
explicitly present in the MF equations for infinite network size.
Our results confirm the MF
prediction of a different behavior of the mean ordering time depending
on whether $\gamma$ is larger or smaller that $5/2$.  In order to check the
predictions of our MF analysis, we report the results of extensive
numerical simulations. These simulations prove that, for symmetric
initial conditions, the dynamics behaves in a way incompatible with
the MF predictions (as already observed numerically in other
nonequilibrium dynamics on complex networks \cite{castellanonMF}),
showing no qualitative differences for $\gamma$ above or below $5/2$, a
value that seems to emerge as an artifact of an invalid MF theory.
Our findings reconcile the results for Glauber dynamics on scale-free
networks with those obtained on other types of structures and calls
for approaches, beyond mean-field, able to capture the observed
phenomenology.

\section{Mean-field theory}
\label{sec:mean-field-theory}
Let us consider an Ising model with spin variables $\sigma=\pm1$ located at
the vertices of a network, which is fully characterized by the
adjacency matrix $A_{ij}$, taking the value $A_{ij}=1$ if the vertices
$i$ and $j$ are connected by an edge, while $A_{ij}=0$ otherwise. The
local field $h_i(t)$ experienced by vertex $i$, and due to the spins
located at its nearest neighbors, is thus given by
\begin{equation}
  h_i(t) = \sum_j A_{ij} \sigma_j(t).
\end{equation}

The Glauber dynamics at zero temperature is defined as follows: At
each time step $t$ we pick at random one of the vertices and we update
the value of its spin according to the local field, that is,
\begin{eqnarray}
  \sigma_i &=& +1  ~~\mathrm{if}  ~~h_i(t) >0\\
  \sigma_i &=& -1  ~~ \mathrm{if} ~~ h_i(t) <0\\
  \sigma_i &=& \pm 1  ~~\mbox{with probability 1/2}  ~~\mathrm{if} ~~ h_i(t)=0.
\end{eqnarray}
After each spin update, time is increased as $t \to t+1/N$, where $N$ is
the number of vertices in the network.

Starting from a disordered uncorrelated configuration, in which each
spin is taken to be positive at random with probability $p$, Glauber
dynamics tends to make neighboring spins equal in sign. Thus, the main
questions raised by this dynamics are whether all spins become equal
in the long run and, if this is the case, how much time this total
ordering requires, and how it depends on the initial fraction $p$ of
positive spins and on the system size $N$.  Particularly interesting
in this respect is the symmetric initial state in which $p=1/2$, since
in this case the dynamics must first break the symmetry of the initial
state in order to reach full order.

To construct a MF theory for the zero temperature Glauber dynamics, we
consider, following Ref.~\cite{Zhou05}, the average dynamical variable
$q_k(t)$, defined as the probability that a vertex of degree $k$ is in
a $+1$ state.  The evolution of the variable $q_k(t)$ depends on the
average local field felt by the vertices of degree $k$, $h_k$. In
terms of this variable, and according to the rules defining the
Glauber dynamics, we can write a rate equation for $q_k(t)$ in the
continuous time approximation (which is valid for infinitely large
network sizes) that takes the form
\begin{eqnarray}
  \frac{d q_k(t)}{d t} &=&  - q_k(t) \mathrm{Prob}[h_k<0] - q_k(t)
  \mathrm{Prob}[h_k=0] \frac{1}{2} \nonumber \\
  &+& (1- q_k(t)) \mathrm{Prob}[h_k>0] 
  + (1- q_k(t)) \mathrm{Prob}[h_k=0] \frac{1}{2},
  \label{eq:8}
\end{eqnarray}
where $\mathrm{Prob}[h_k > 0]$, $\mathrm{Prob}[h_k<0]$, and
$\mathrm{Prob}[h_k=0]$ are the normalized probabilities that the local
field at the vertices of degree $k$ is positive, negative or zero,
respectively.  Rearranging the terms in Eq.~(\ref{eq:8}) we can write
\begin{equation}
  \frac{d q_k(t)}{d t} =  - q_k(t) + \mathrm{Prob}[h_k>0] +
  \frac{1}{2}\mathrm{Prob}[h_k=0].
  \label{eq:3}
\end{equation}

In order to estimate the probabilities of the local field $h_k$, the
relevant quantity to consider is the probability $Q_k$ that an edge
departing from a vertex of degree $k$ points to a $+1$ spin. For a
generic network, statistically characterized by its degree distribution
$P(k)$ and its degree correlations given by the conditional
probability $P(k'|k)$ that a vertex of degree $k$ is connected to a
vertex of degree $k'$ \cite{alexei}, the probability that an edge from a
vertex of degree $k$ points to a $+1$ spin is given by
\begin{equation}
  Q_k = \sum_{k'} P(k'|k) q_k,
\end{equation}
that is, it is proportional to the probability that a vertex $k$ is
connected to a vertex of degree $k'$ times the probability that this
vertex is in a $+1$ state, averaged over all possible values of the
degree $k'$. For random uncorrelated networks, the conditional
probability takes the simplified form $P(k'|k) = k P(k) / \av{k}$, and
therefore the probability $Q_k$ is independent of $k$ and can be
written as~\cite{Zhou05}
\begin{equation}
  Q = \frac{1}{\av{k}} \sum_k k P(k) q_k.
  \label{eq:2}
\end{equation}
Assuming that the probability of having a positive spin at the end of
an edge departing from a $k$ vertex is independent of the values of
the spins at the extremes of the other $k-1$ edges (which corresponds
to a MF assumption), we have that the probability that $\ell$ edges from
a $k$ vertex point to positive spins is given by a binomial
distribution.
The local field at a $k$ vertex can be zero only if $k$ is even and
exactly half of its edges point to $+1$ spins. Thus
\begin{equation}
  \mathrm{Prob}[h_k=0] =
\binom{k}{k/2} Q^{k/2} (1-Q)^{k/2},
\end{equation}
for $k$ even.
On the other hand, $h_k$ is positive when more than half of its edges
point to $+1$ spins. Therefore, the probability of observing a
positive local field is 
\begin{equation}
\mathrm{Prob}[h_k>0] =
\sum_{\ell = \lceil(k+1)/2\rceil }^k \binom{k}{\ell} Q^\ell (1-Q)^{k-\ell},
\end{equation}
where $\lceil x\rceil$ is the smallest integer larger than or equal to $x$. In
this way, we can write the rate equation for the probabilities $q_k$
as
\begin{equation}
  \frac{d q_k(t)}{d t} =  - q_k(t) + \Phi_k(Q),
\label{eq:1}
\end{equation}
where~\cite{Zhou05}
\begin{equation}
  \Phi_k(Q) = \sum_{\ell = \lceil k/2\rceil}^k \left[1-\frac{1}{2}\delta_{\ell, k/2}\right] 
  \binom{k}{\ell} Q^\ell (1-Q)^{k-\ell}.   
\label{phi_k}
\end{equation}
By introducing Eq.~(\ref{eq:1}) into the definition Eq.~(\ref{eq:2}) we
obtain a closed rate equation for the quantity $Q$, namely 
\begin{equation}
   \dot Q
= - Q + \Psi(Q),
\end{equation}
with
\begin{equation}
  \Psi(Q) = \frac{1}{\av{k}} \sum_k k P(k) \Phi_k(Q).
\end{equation}

It is useful to write the dynamics in terms of the new variable $y =
Q-1/2$. In the initial state, where all spins point up or down with
probability $p$ and $1-p$, respectively, we have $Q(t=0)=p$ and
therefore $y(t=0)=p-1/2$. In the symmetric disordered state, with
$p=1/2$, we have $y(t=0)=0$, while the ordered states with all spins
up or down correspond to $y=1/2$ and $y=-1/2$, respectively. Thus, $y$
plays the role of a convenient order parameter.  The rate equation for
the quantity $y$ can be written as
\begin{equation}
\dot y= -y -\frac{1}{2} + \psi(y),
\label{eqpsi}
\end{equation}
where we have defined the new function
\begin{equation}
\psi(y) \equiv \Psi\left(\frac{1}{2}+y \right)= 
\sum_k \frac{k P(k)}{\av{k}} \Phi_k\left(\frac{1}{2}+y\right).
\label{eq:9}
\end{equation}
The function $\psi(y)$ is equal to $1/2$ for $y=0$ and goes symmetrically
to 1 [0] for $y = 1/2~[y=-1/2]$. In terms of this variable, we can
study the time that it takes for the system to become ordered by
defining the ordering time $t_{ord}$ such that $|y(t_{ord})| = y_F$,
where $y_F$ is an arbitrary chosen positive value such that $|p-1/2|< y_F<1/2$.

>From the explicit evaluation of $\psi(y)$, presented in the Appendix, it
turns out that, for an uncorrelated network with degree distribution
$P(k) \sim k^{-\gamma}$, there are different limiting regimes for
different values of $y$ and the degree exponent $\gamma$.
For simplicity we consider only the case $y>0$;
analogous considerations apply in the symmetric case $y<0$.

$\mathbf{2<\gamma<5/2}$: The behavior of $\psi(y)$ can be approximated,
depending on the value of $y$ and of the minimum degree $k_0$ and
degree cut-off $k_c$ of the network in the following form:
\begin{equation}
  \begin{array}{ll}
    \psi(y) -1/2 \sim c_2(\gamma) k_0^{\gamma-2}k_c^{5/2-\gamma} y, & y<y_c\\
    \psi(y) -1/2 \sim c_3(\gamma) k_0^{\gamma-2} y^{2(\gamma-2)},  
& y_c < y < y_{NL}\\
    \psi(y) -1/2 \sim 1/2,  & y > y_{NL}\\
  \end{array}
  \label{eq:14}
\end{equation}
where $y_c =k_c^{-1/2}$ and $y_{NL} = k_0^{-1/2}$.

Inserting these forms into the dynamic
equation Eq.~(\ref{eqpsi}), we obtain the temporal evolution of the
order parameter $y$. For $y<y_c$, there is a regime
with $y$ growing exponentially with time, i.e.
\begin{equation}
y(t) = y_T \exp[(c_2(\gamma) k_0^{\gamma-2} k_c^{5/2-\gamma} -1) (t-t_T)],
\label{eq:5}
\end{equation}
with an exponential factor that depends on the network cut-off $k_c$.
Here we have denoted as $t_T$ and $y_T$ the values of $t$ and $y$ at
the beginning of this regime.  For reasons that will be clear below,
these may or may not coincide with the initial values $t=0$ and
$y(0)$.  For $y_c < y < y_{NL}$, on the other hand, the order parameter grows
instead as a power-law,
\begin{equation}
  y(t) = \left[y_c^{1-2(\gamma-2)} +
    c_3(\gamma) k_0^{\gamma-2}(t-t_c)\right]^{1/[1-2(\gamma-2)]},
  \label{eq:15}
\end{equation}
where $t_c$ is the time at which the order parameter reaches the value
$y_c$.  Finally, for $y > y_{NL}$, $\psi(y)$ is approximately equal to
$1$ and the dynamic equation is $\dot y = 1/2-y$, so that
\begin{equation}
y(t) =
  \frac{1}{2} + \left(y_{NL} - \frac{1}{2} \right) e^{-(t-t_{NL})},
  \label{eq:4}
\end{equation}
where $t_{NL}$ is the time at which the value $y_{NL}$ is reached,
that is, $y(t_{NL})= y_{NL}$.

$\mathbf{\gamma > 5/2}$: In this case $y_c$ does not play any role;
the function $\psi(y)$ is independent
of $k_c$ and it can be approximated for $y < y_{NL}$ as
\begin{equation}
  \psi(y) \sim  1/2 + c_1(\gamma) k_0^{1/2} y.
  \label{eq:16}
\end{equation}
The behavior $y$ is thus again exponential, but now with a
prefactor independent of $k_c$, namely,
\begin{equation}
  y(t) = y_T \exp[(c_1(\gamma) k_0^{1/2}  -1) (t-t_T)],
\label{eq:17}
\end{equation}
For $y > y_{NL}$, $\psi(y)$ the scenario is the same of the
previous case, with an exponential approach to the asymptotic
value $y=1/2$.

As a consequence of the previous expressions, the total ordering time
has, in the most general case, four contributions, namely,
\begin{equation}
t_{ord} = t_T + \Delta t_c + \Delta t_{NL}+\Delta t_{ord},
\label{t_ord}
\end{equation}
where $\Delta t_c = t_c- t_T$, $\Delta t_{NL} = t_{NL}- t_c$, 
and $\Delta t_{ord} = t_{ord}-t_{NL}$.

The simplest term is the last one, $\Delta t_{ord}$,
that, from Eq.~(\ref{eq:4}), is given by
\be
\Delta t_{ord} \approx \ln \left[\frac{1 - 2 y_{NL}}{1- 2 y_F} \right].
\ee
It is independent of $N$, $p$, and, in practice, also of $k_0$.
The other terms depend instead on the value of $\gamma$.

For $\gamma > 5/2$, $\Delta t_c$ and $\Delta t_{NL}$ are merged
in a single term, which is, from Eq.~(\ref{eq:17}),
\begin{equation}
  \Delta t_c + \Delta t_{NL} \approx k_0^{-1/2} \ln \left[\frac{y_{NL}}{y_{T}} \right].
  \label{t_L}
\end{equation}

For $2<\gamma< 5/2$, instead, we obtain from Eq.~(\ref{eq:15})
\begin{equation}
\Delta t_{NL} \approx \frac{y_{NL}^{1-2(\gamma-2)} - 
y_c^{1-2 (\gamma-2)}}{k_0^{\gamma-2}},
\end{equation}
while from Eq.~(\ref{eq:5}) we have
\begin{equation}
  \Delta t_c \approx
  k_0^{2-\gamma } k_c^{\gamma-5/2} \ln \left[\frac{y_c}{y_{T}} \right].
\label{eq:6}
\end{equation}

Let us now discuss the additional term $t_T$.  The deterministic
equation of motion~(\ref{eqpsi}) holds strictly only in the limit of
infinite network size. For finite $N$ an additional random term
appears, giving a diffusive contribution to the dynamics of $y$ that
competes with the deterministic drift described by Eq.~(\ref{eqpsi}).
As it will be shown below, this contribution vanishes as $N$ diverges,
so that it can be generally neglected with respect to the drift, {\it
  unless} $y(0)=0$, i.e., for symmetric initial conditions.  In such a
case also the drift vanishes, and the point $y=0$ is an unstable
equilibrium point, that can be characterized by a potential $V(y) =
-\int^y ( \psi(y) -y -1/2) d y$ with a parabolic shape for very small $y$.
Under the deterministic MF dynamics a system placed at $y=0$ will stay
there forever, and the diffusive term is thus needed to provide the
perturbation necessary to take the system out of the initial
equilibrium point.  The interplay between drift and diffusion
determines the time $t_T$ at which drift starts to dominate.

The origin of this diffusive term is easy to understand: As discussed
at the beginning of this Section, the Glauber dynamics at zero
temperature is defined by means of a sequential updating, in which at
each time interval $\Delta t = 1/N$ a single spin is flipped, causing an
average increment of the dynamic variable $\Delta Q = 1/N$.  The initial
condition $y=0$, $Q=1/2$, corresponds to a state in which every site
is randomly assigned a $+1$ or a $-1$ spin with probability $p=1/2$.
Therefore, the initial evolution is dominated by a random flipping of
spins, driven by the value of the local field $h_i$, that induces a
diffusive motion of the variable $y$ until a nonzero value is reached,
large enough to drive the dynamics out of the unstable equilibrium
point. To be more concrete, when $y \simeq 0$, or $Q \simeq 1/2$, the drift term
appearing in the equation of motion is very small.  In particular, the
deterministic drift velocity is given by
\begin{equation}
v = (1-Q) \Psi(Q) - Q [1 - \Psi(Q)] = \psi(y)-1/2-y.
\end{equation}
The diffusive component, on the other hand, has a diffusive constant
that can be estimated as
\begin{equation}
D = (\Delta Q)^2/(2 \Delta t) = 1/(2N).
\end{equation}
For small values of $y$, the drift grows with $y$, while diffusion is
independent of it.  Starting with $y(0) \simeq 0$, for small $|y|$
diffusion prevails and the drift can be neglected.  The diffusion
leads to increasing fluctuations, and after some time a fluctuation
leads to a threshold value $|y_T|$ sufficiently large for the bias to
start prevailing, and drive the variable far from the equilibrium
point.  From this time on, bias dominates and the diffusive term in the
equation of motion becomes negligible.

The value of $y_T$ can be determined as follows: When the system is in
the position $y$, the time $t_D$ needed to go back diffusively to 0 is
given by $|y| = \sqrt{D t_D}$.  During this time interval the bias
will induce a drift $|v(y)| t_D = |v(y)| y^2/D$.  If the drift is
smaller than $|y|$ the system is able to go back to 0, and diffusion
dominates. If the drift is larger than $|y|$ then bias dominates. The
condition that defines  $y_T$ is then 
\begin{equation}
v(y_T) \frac{y_T^2}{D} = |y_T|.
\label{eqv}
\end{equation}
Inserting Eq.~(\ref{eq:16}) into Eq.~(\ref{eqv}) we
obtain, for $\gamma>5/2$,
\be 
|y_T| \approx  N^{-1/2} k_0^{-1/4}, 
\ee 
while the time spent diffusing is
\be
t_T \approx \frac{y_T^2}{D} \approx k_0^{-1/2}.
\ee
Notice that this time does not depend on $N$.
On the other hand, for $2<\gamma<5/2$, we obtain, using the scaling of the
cut-off with the network size, given by  $k_c \sim N^{1/2}$ for
uncorrelated scale-free networks \cite{mariancutofss},
\be 
|y_T| \approx k_0^{(2-\gamma)/2} N^{(\gamma-9/2)/4},
\ee
and 
\be
t_T \approx  k_0^{(2-\gamma)} N^{(\gamma-5/2)/2},
\ee
a time that goes to zero when increasing $N$. We are now in the
position to summarize the MF estimates for the 
ordering time.

For asymmetric initial conditions ($p \neq 1/2$, $y(0)>0$) we can set
$t_T=0$ and $y_T=y(0)$. In this case, we find at leading order
for $\gamma>5/2$
\begin{equation}
t_{ord} \approx 
\mbox{const} + k_0^{-1/2} \ln y(0),
\label{t_asym}
\end{equation}
while for $2 < \gamma < 5/2$ we obtain
\begin{equation}
  t_{ord}  \approx \mbox{const} 
  + k_0^{2-\gamma} N^{(\gamma-5/2)/2} \ln \left[\frac{N }{y(0)} \right],
\label{t_asym2}
\end{equation}
As we can observe, for finite $y(0)$ the ordering time is finite
in the infinite network size limit. For
finite networks, it exhibits a decreasing size correction in the case
$2 < \gamma < 5/2$.

\begin{figure}

  \centerline{\epsfig{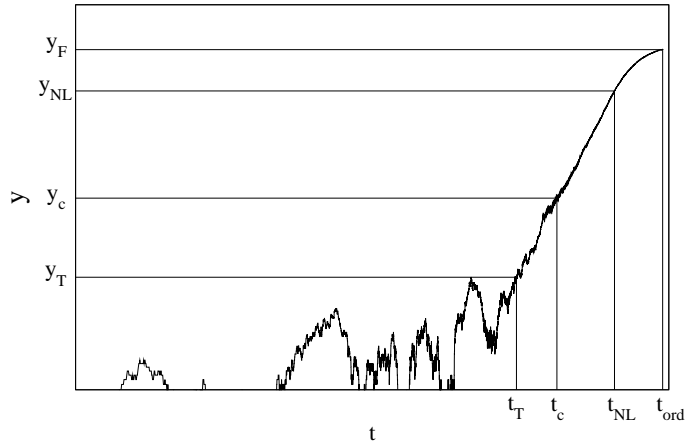}}
  \caption{Sketch of the behavior of the evolution of $y(t)$ as a function
    of time for $2<\gamma<5/2$. Both axes are on logarithmic scale.}
  \label{sketch}
\end{figure}

For symmetric initial conditions, $y(0)=0$, and $\gamma>5/2$ we have, at
leading order,
\begin{equation}
t_{ord} \approx  \mbox{const} +  k_0^{-1/2} \ln N ,
\label{eq:7}
\end{equation}
that is, the ordering time diverges logarithmically with the network
size $N$.  For $2<\gamma<5/2$, on the other hand, we obtain 
\begin{equation}
  t_{ord} \approx \mbox{const}  + 
  k_0^{2-\gamma}
  N^{(\gamma-5/2)/2} \ln N.
\label{eq:7_2}
\end{equation}
The ordering time is now a decreasing function of $N$ that tends to a
constant value in the limit  $N \to\infty$.
For illustration purposes a sketch of the different temporal
scales and the related behaviors of $y$ is depicted in Fig.~\ref{sketch}.

\section{Numerical results}

In this Section we compare the theoretical results derived above
within the MF formalism with extensive numerical simulations of the
Glauber dynamics at zero temperature, performed on uncorrelated
networks with a power-law degree distribution.  The networks used have
been generated using the uncorrelated configuration model (UCM)
defined in Ref.~\cite{ucmmodel}. The network is built as follows: We
start by assigning to each vertex $i$ in a set of $N$ initially
disconnected vertices a random degree $k_i$, extracted from the
probability distribution $P(k)$, subject to the constraint $k_0 \leq k_i
\leq k_c = N^{1/2}$ and $\sum_i k_i$ even.
Afterward, the network is constructed by
randomly connecting the vertices with $\sum_i k_i /2$ edges, respecting
the preassigned degrees and avoiding multiple and self-connections
between vertices. Using this algorithm, it is possible to create
random networks with any preassigned degree distribution. For the case
of networks with a degree distribution given by a power law, $P(k) \sim
k^{-\gamma}$, the restriction $k \leq N^{1/2}$ for any degree exponent
guarantees that the networks generated in this way are completely
uncorrelated \cite{mariancutofss}.  Notice that if such a restriction
is not enforced, correlations unavoidably arise for
$\gamma<3$~\cite{mariancutofss}.

We start considering the case $\gamma=4$ as a representative of networks
with $\gamma>3$. In this case the connectivity pattern does not show
strong degree fluctuations ($\langle k^2 \rangle < \infty$ for $N\to\infty$) and one does not
expect to observe the special effects induced by the presence of hubs,
relevant for other dynamical processes on networks.  MF theory
predicts the ordering time to be given by Eq.~(\ref{t_asym}) for
$y(0)>0$ and by Eq.~(\ref{eq:7}) for $y(0)=0$.  However, even in this
simple case, we find that the observed phenomenology is in striking
disagreement with the expected theoretical results.

\begin{figure}
  \centerline{\epsfig{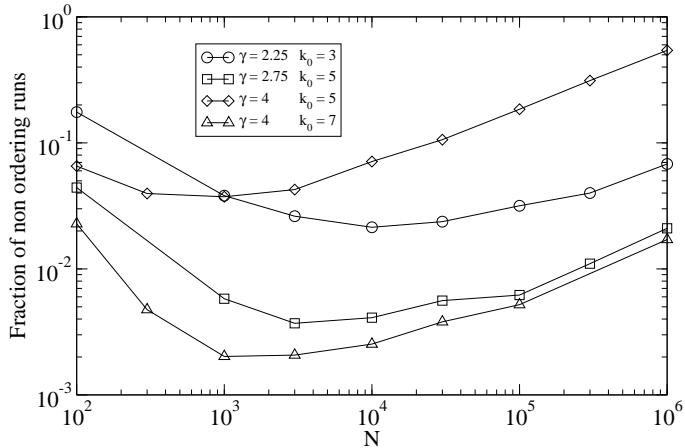}}
  \caption{Plot of the probability that a run starting from the
    symmetric initial state $y(0)=0$ does not lead to complete order
    as a function of $N$, for networks with various values of $\gamma$ and
    $k_0$.}
  \label{p_dis}
\end{figure}

The first, fundamental, discrepancy lies in the very existence of
order in the system for large times, At the MF level, all instances of
the dynamics order sooner or later.  From the numerical evidence,
however, we conclude that networks starting from a symmetric initial
state do not get ordered in the large $N$ limit.  To reach this
conclusion, we have performed a large number of runs for several
values of $N$, starting from  $y(0)=0$. We observe that some of
them do not lead to a fully ordered state, remaining instead trapped
in some (meta)stable states that last forever.  Fig.~\ref{p_dis}
reports the fraction of these runs that do not lead to complete
ordering.  After a minimum for a network size that apparently depends
on the value of $k_0$, this fraction grows with $N$.  Extrapolating
for large $N$, we conclude that an infinite network never gets
ordered. This absence of ordering of the Glauber dynamics a $T=0$ on
complex networks had already been pointed out in
Ref.~\cite{Castellano05}.  Fig.~\ref{p_dis} demonstrates also that the
lack of ordering occurs for all values of $\gamma$, either larger or
smaller that $5/2$.  From this figure we can also observe that the
fraction of nonordering runs decreases with increasing $k_0$, in such
a way that for very large values of $k_0$ and moderate values of $N$,
this fraction is apparently zero (i.e.  all realizations get ordered).
This fact, however, is a trivial consequence of the nonordering
probability becoming too small for the number of runs that can be
performed. In Ref.~\cite{Zhou05}, where large values of $k_0$ were
used, this absence of ordering was not noticed.

\begin{figure}
  \centerline{\epsfig{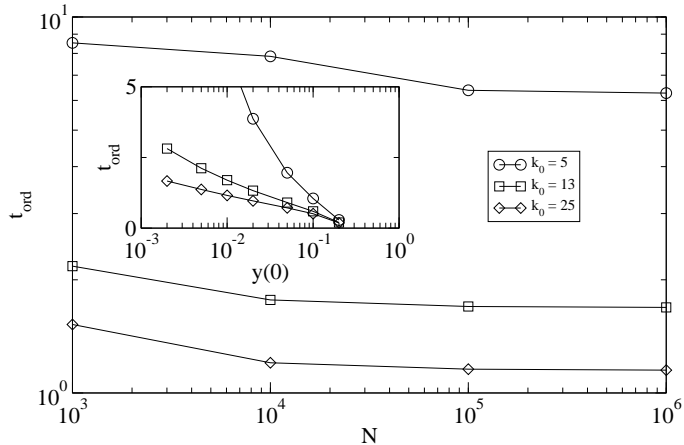}}
  \caption{Average value of the ordering time $t_{ord}$,
    for $\gamma=4$ as a function of the network size, for fixed $y(0)=0.01$
    and different values of $k_0$. Inset: Asymptotic ordering time as
    a function of the initial $y(0)$.}
  \label{time}
\end{figure}

In view of this result, the very notion of an ordering time becomes
somewhat fuzzy, in the sense that an unrestricted average over all runs
will yield an infinite ordering time. For practical purposes,
nevertheless, we can define an operational ordering time by performing
an average over the restricted ensemble of those runs that actually
order. Thus we have computed the average ordering time, defined as the
average time $t_{ord}$ needed for the ordering parameter $y$ to reach
the arbitrary value $|y_F|=1/4$ \cite{Zhou05}.  In Fig.~\ref{time} we
report the values of $t_{ord}$ obtained for networks with $\gamma=4$,
starting from nonsymmetric initial conditions $y(0)>0$ and different
values of $k_0$, as a function of the network size $N$. As predicted
from the MF result Eq.~(\ref{t_asym}), we observe that $t_{ord}$
becomes independent of $N$, for sufficiently large network sizes. In
the inset of Fig.~\ref{time} we plot the asymptotic value of $t_{ord}$
for the largest network size considered $N=10^6$, as a function of
$y(0)$. Here we can observe that the MF prediction $t_{ord} \sim - \ln
y(0)$, Eq.~(\ref{t_asym}), seems to be fulfilled, at least for
sufficiently large values of $k_0$ and $y(0)$. For small values of
$y(0)$ and $k_0$ the convergence of $t_{ord}$ with $N$ is rather slow,
a fact that hides the asymptotic logarithmic behavior.

\begin{figure}
  \centerline{\epsfig{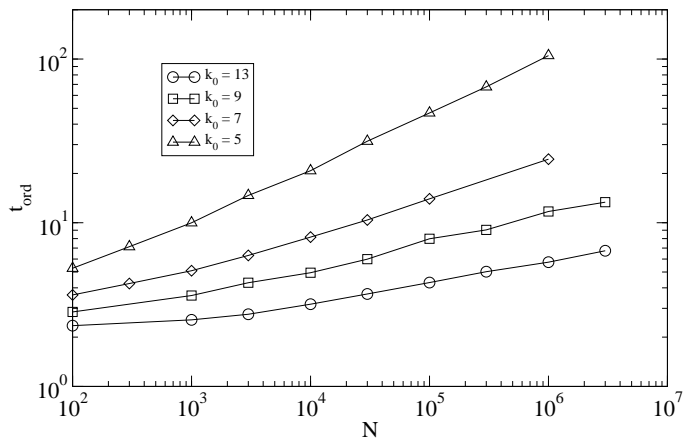}}
  \caption{Average value of the ordering time $t_{ord}$,
    for $\gamma=4$ as a function of the network size, for symmetric initial
    conditions $y(0)=0$ and several values of the minimum degree
    $k_0$.}
  \label{timesym}
\end{figure}

In Fig.~\ref{timesym} we show the results obtained for the ordering
time in networks with $\gamma=4$ and symmetric initial conditions $y(0)=0$,
for different values of the minimum degree $k_0$. As we can see from
this plot, the average ordering time $t_{ord}$ seems to grow with $N$
as a power-law, with effective exponents depending on the particular
value of $k_0$. For large values of $k_0$, the
exponent is so small that the behavior could be confused with a
logarithmic one. However, for small $k_0$, the power-law growth is quite
evident, in strong disagreement with the MF prediction,
Eq.~(\ref{eq:7}).

\begin{figure}
 \centerline{\epsfig{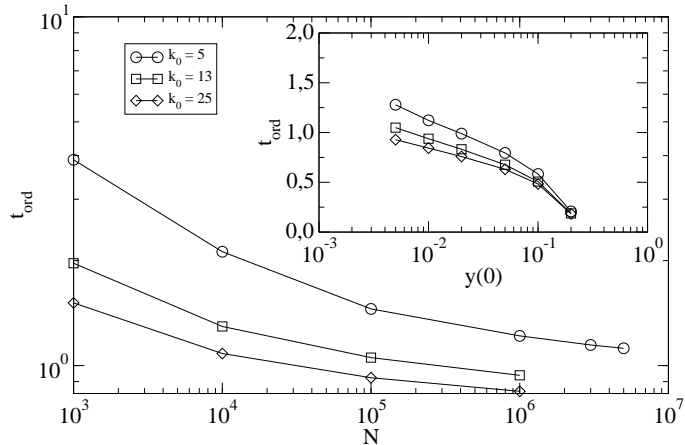}}
  \caption{Average value of the ordering time $t_{ord}$,
    for $\gamma=2.25$ as a function of the network size,
    for fixed $y(0)=0.01$ and different values of $k_0$.
    Inset: Asymptotic ordering time as a function of the initial $y(0)$.}
  \label{asymmg=2.25}
\end{figure}

We have performed this same kind of analysis for two other values of
the degree exponent, namely $\gamma=2.75$ and $\gamma=2.25$, in order to check
the possible effects of a strongly inhomogeneous degree distribution,
with diverging second moment, and also to asses whether the values of
$\gamma$ smaller than $5/2$ affect the behavior of the ordering time, as
predicted by MF theory.  Results for asymmetric initial conditions
$y(0)>0$ for $\gamma=2.25$ are reported in Fig.~\ref{asymmg=2.25}. In this
case, we find that the convergence to the asymptotic value of the
ordering time for large $N$ is extremely slow.  As a consequence, it
is not possible to make a definite statement about the validity of MF
based on the present numerical evidence.  For the largest $N$ that
could be reached, the behavior of $t_{ord}$ for small $y(0)$ does not
seem to be logarithmic, inset in Fig.~\ref{asymmg=2.25}, in apparent
contradiction with Eq.~(\ref{t_asym2}).  However, it is possible that
this discrepancy could disappear if larger values of $N$ could be
considered.

\begin{figure}
  \centerline{\epsfig{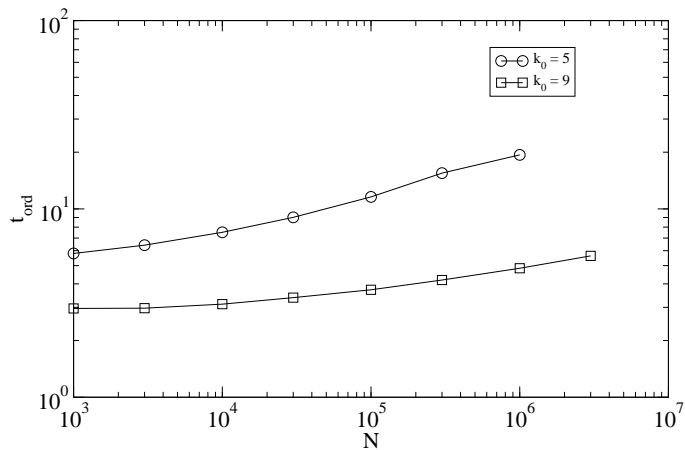}}
  \caption{Average value of the ordering time $t_{ord}$,
    for $\gamma=2.75$, as a function of the network size,
    for symmetric initial conditions $y(0)=0$ and two values
    of the minimum degree $k_0$.}
  \label{time_g2.75}
\end{figure}

\begin{figure}
  \centerline{\epsfig{file=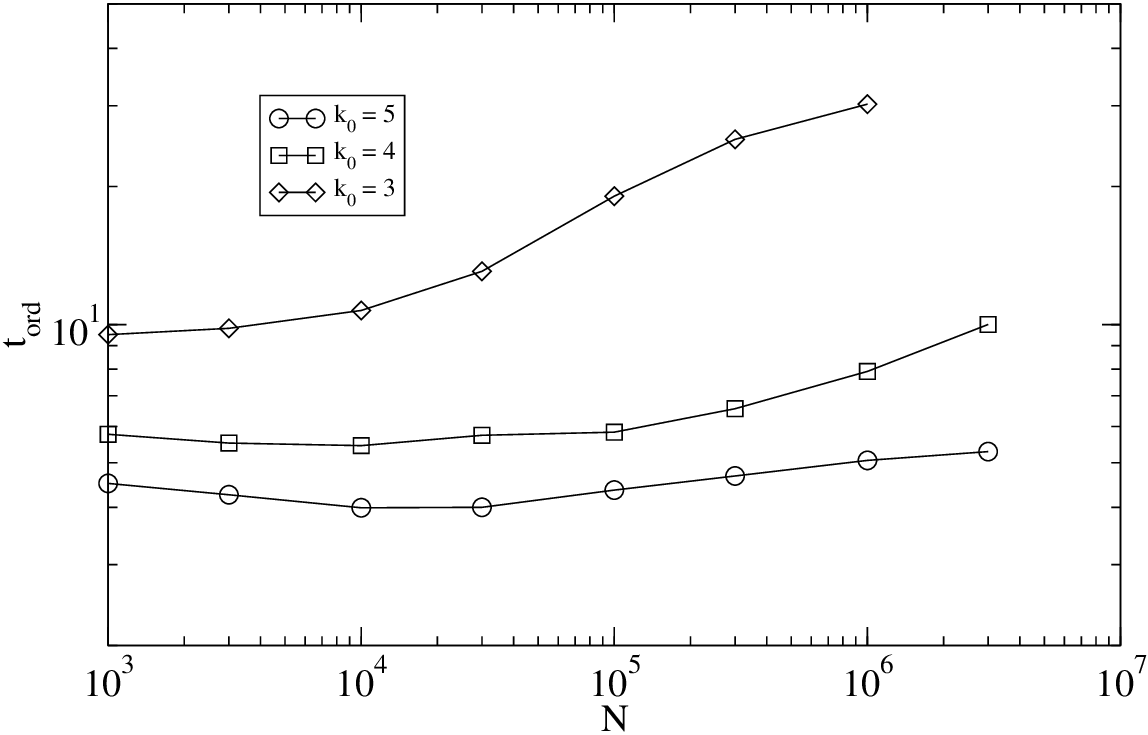, width=9cm}}
  \caption{Average value of the ordering time $t_{ord}$,
    for $\gamma=2.25$, as a function of the network size,
    for symmetric initial conditions $y(0)=0$ and three values
    of the minimum degree $k_0$.}
  \label{time_g2.25}
\end{figure}

Much clearer is the situation for the symmetric initial condition
$y(0)=0$: the behavior is qualitatively the same (and different from
MF) independently of the value of $\gamma$.  We have already seen in
Fig.~\ref{p_dis} that the absence of ordering for large $N$ holds for
any value of $\gamma$. But the similarity is stronger  when the ordering
time $t_{ord}$ is measured as a function of $N$.  As displayed in
Figs.~\ref{time_g2.75} and~\ref{time_g2.25}, $t_{ord}$ diverges at
large $N$ as a power-law, as opposed to the logarithmic growth and the
constant behavior predicted by MF for $\gamma=2.75$ and $\gamma=2.25$,
respectively.

The conclusion of this numerical analysis is that MF theory is unable
to account for the behavior of the ordering time of the Glauber
dynamics at zero temperature with symmetric initial conditions. The
physical reason for this failure can be understood by considering in
more in detail the dynamical evolution of the system.

The symmetric initial state is created by randomly placing $+1$ and
$-1$ spins on the network with probability $p=1/2$.  This procedure is
akin to a percolation process~\cite{stauffer94} of both the positive
and negative spins with the same probability $p$. In random networks,
the percolation threshold is given by~\cite{havlin00}
\begin{equation}
  p_c = \frac{1}{\frac{\fluck{k}}{\av{k}}-1}.
  \label{eq:10}
\end{equation}
In scale-free networks with $\gamma<3$, the degree fluctuations diverge,
$\fluck{k} \to \infty$, and $p_c \to 0$ in the limit of large network sizes.
In the case of homogeneous networks, with finite $\fluck{k}$, we
observe that $p_c \sim 1/ \av{k}$, and thus it becomes small for large
average degree or $k_0$. Hence for reasonable values of $k_0$, all
networks exhibit a very small percolation threshold.  Therefore,
symmetric initial conditions give rise to two giant components, $G_+$
and $G_-$, with approximate size $N/2$, each one corresponding to a
connected set of vertices occupied by $+1$ or $-1$ spins,
respectively.
\begin{figure}
  \centerline{\epsfig{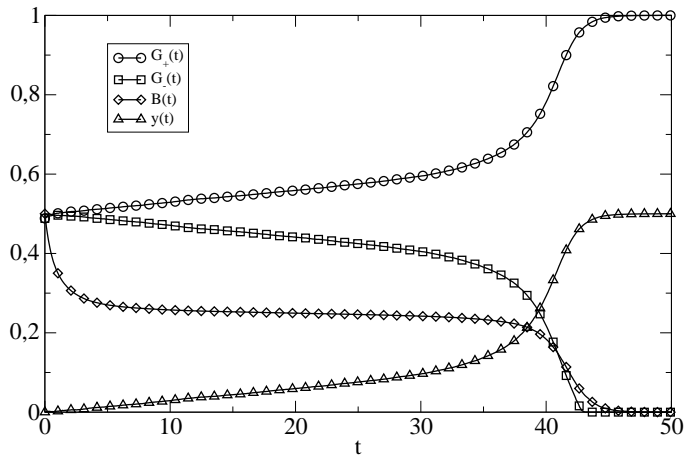}}
  \caption{Temporal evolution of the size of the giant components ($G_+$ and
    $G_-$), the size of the boundary $B$ and the order parameter $y$
    for a run on a network with $N = 10^5$, $\gamma=4$, $k_0=5$ with
    symmetric initial conditions.}
  \label{giants}
\end{figure}
In Fig.~\ref{giants} we plot, as a function of time and in a single
run, the relative size of the giant components of $+1$ spins,
$G_+(t)$, and $-1$ spins, $G_-(t)$, and the relative size of the
boundary $B(t)$, defined as the fraction of edges with spins of
different sign at the ends. For illustration, we also plot the value
of the order parameter $y(t)$ corresponding to that particular
ordering run.  From this Figure we can observe that the dynamics is
governed by the competition of the giant components of different
signs, the $G_+$ growing at the expense of the $G_-$. It is noteworthy
that during most of the evolution the boundary occupies a constant and
sizeable fraction of all the edges.  In the case shown in the Figure
the competition ends abruptly with the positive giant component $G_+$
invading the whole system.  In some cases instead, the competition
ends up in a stalemate: the system reaches a stationary configuration
with the two coexisting giant components basically frozen with only
some spins of even degree freely flipping back and forth at the
boundary.

\begin{figure}
  \centerline{\epsfig{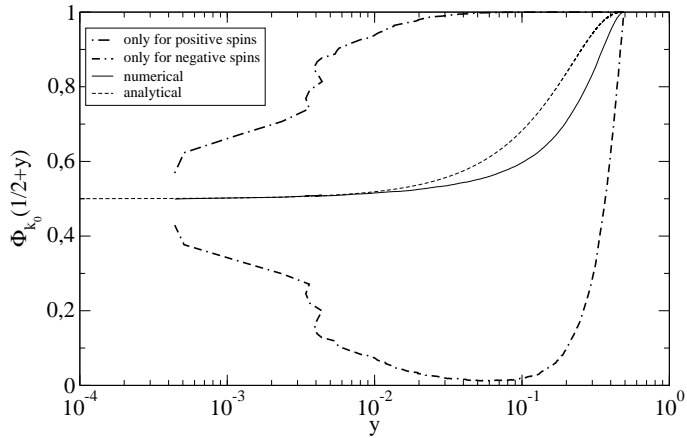}}
  \caption{Plot of the probability that the local field of a vertex
with $k_0=5$ links is positive. Data are for a system with $\gamma=4$
and $N=100000$. The meaning of the different curves is explained in
the text.}
  \label{psi}
\end{figure}

In the initial state $B(t=0)=2p(1-p)$. After a short transient $B(t)$
reaches the plateau. During this short time interval strong
correlations build up in the system, in the sense that $+1$ spins
become surrounded with high probability with a majority of positive
spins, while most neighbors of a $-1$ spin are almost certainly
negative. This fact can be quantitatively measured by computing the
mean probability that a vertex with a given spin $\sigma$ has a positive
local field $h$.  This kind of measurement is reported in
Fig.~\ref{psi} for homogeneous networks with degree exponent $\gamma=4$,
where we display the probability that the local field of a site of
degree $k_0$ is positive. Within the mean-field treatment, this
quantity is given by the function $\Phi_{k_0}(1/2+y)$, Eq.~(\ref{phi_k}),
plotted as a dashed line.
The solid line in the middle is the numerical evaluation of the same
quantity.  The other two  lines are instead numerical evaluations
of the same probability, but averaged only over all sites with $s_i =
1$ (top) and $s_i = -1$ (bottom).  The middle solid line (that is the
average of the top and bottom lines) is quite similar to the
analytical result, but the other two are completely different.  The
mean-field treatment assumes that the local field is completely
uncorrelated from the value of the spin, and depends at most on the
degree of the vertex in which the spin is located.  In the real
system, however, during virtually all the dynamics, a positive spin has
almost surely a positive local field, while a negative spin has a
negative field.  These strong correlations, completely missed by the
MF approach, are at the root of the nontrivial behavior shown by the
zero temperature Glauber dynamics.

\section{Conclusions}
In this paper we have investigated the ordering process of the zero
temperature Glauber dynamics when Ising spins are placed on the
vertices of an uncorrelated network with degree distribution
decreasing as a power-law of exponent $\gamma$.  We have first developed a
mean-field theory for the problem, expanding and improving on the earlier
approach presented in Ref.~\cite{Zhou05}. Within the MF theory we
identify a suitable order parameter and compute in detail the time
needed for the system to order.  While for asymmetric initial conditions
the ordering time is finite, for symmetric initial conditions there is a
transition: $t_{ord}$ is finite for $\gamma<5/2$, while it diverges
logarithmically with the system size $N$ for $\gamma>5/2$.

The validity of these analytical results has been checked by means of
extensive numerical simulations. While for asymmetric initial
conditions we find an apparently reasonable agreement with MF theory,
when the probability of a positive initial spin is $p=1/2$, the
behavior is at odds with the MF results and highly nontrivial.  In
particular, it turns out that there is a finite probability for a run
to get trapped forever in a stationary disordered state. This
probability grows with $N$ so that complete ordering is not reached
for asymptotically large $N$.  This conclusion, in striking
disagreement with the MF predictions, holds for all values of $\gamma$.  If
we compute the ordering time restricted only to those runs that
actually lead to full order, other unexpected results arise. For large
$N$ and independently of the degree exponent, $t_{ord}$ appears to
diverge as a power-law of the system size, the effective exponent
being a function of $\gamma$ and the smallest degree $k_0$. One of the main
conclusions of our work is, therefore, that mean-field theory does not
provide a correct description of the ordering dynamics of the Glauber
model at zero temperature. In this sense, the peculiar exponent
$\gamma_c=5/2$ turns out to be an artifact produced by a MF theory that is
not capable to describe the system, and it does not bear apparently
any relevance for the true behavior of the Glauber dynamics.

The failure of the MF theory for symmetric initial conditions is
reminiscent of similar observations made on other dynamics on
scale-free networks \cite{castellanonMF}.  In the present case, this
failure can be traced back to the breakdown of the assumption that the
local field of a spin is independent of its value. This breakdown is
explicitly shown in the formation of two giant components of positive
and negative spins, competing through an extensive boundary, which
induces the presence of strong correlations.

It is important to remark that the nontrivial phenomenology uncovered
in this paper is not a consequence of the scale-free property of the
network, since it appears already for $\gamma=4$.  The nontriviality of
this problem is thus totally unrelated from the heterogeneous or
homogeneous nature of the substrate.  Somewhat surprisingly, the
behavior is qualitatively the same for any $\gamma>2$: In this case a
scale-free topology seems not to have any relevant effects.  The key
to a deeper understanding of this problem lies likely in a detailed
analysis of the properties of the two giant clusters competing in the
system and of the boundary separating them.  Along with an
investigation of the effect of temperature in the problem, this
remains as an interesting open issue for future work.

\section*{Acknowledgments}
  We thank Miguel Angel Mu\~noz for useful discussions.
  R. P.-S. acknowledges financial support from the Spanish MEC (FEDER),
  under project No.  FIS2004-05923-C02-01 and additional support from
  the Departament d'Universitats, Recerca i Societat de la Informaci\'o,
  Generalitat de Catalunya (Spain). C. C. acknowledges financial support from
  Azione Integrata Italia-Spagna IT1792/2004.

\appendix
\section{}
In this Appendix we  derive analytically the asymptotic behavior of the
function
\begin{equation}
\psi(y) = \sum_k \frac{k P(k)}{\av{k}} \Phi_k\left(\frac{1}{2}
  +y\right) 
\end{equation}
for finite networks, in which the the degree $k$ is restricted to the
range $k \in [k_0, k_c]$, $k_0$ being the smallest degree in the network
and $k_c$ the degree cut-off or maximum degree present in the
network. In the continuous degree approximation, the degree
distribution takes in this case the form 
\begin{equation}
  P(k) = \frac{(\gamma-1) k_0^{\gamma-1}}{1 - (k_c/k_0)^{1-\gamma}} k^{-\gamma}.
\end{equation}

The function $\Phi_k(1/2+y)$ has the generic form, neglecting the
Kronecker symbol,
\begin{equation}
   \Phi_k(1/2+y) = \sum_{l=k/2}^k p_{k,l}(y),
   \label{eq:21}
\end{equation}
where
\begin{equation}
  p_{k,l}(y) = \binom{k}{l} (1/2+ y)^l (1/2-y)^{k-l}.
  \label{eq:13}  
\end{equation}
For large values of $k$, the binomial distribution Eq.~(\ref{eq:13})
can be approximated by a normal distribution,
\begin{equation}
  p_{k,l}(y) \simeq \frac{1}{\sigma \sqrt{2 \pi }}
 \exp\left(-(l-m)^2/2 \sigma^2\right),
\end{equation}
with
\begin{eqnarray}
  m &=& \sum_l l p_{k,l}(y) = k \left(1/2 + y \right), \\
  \sigma^2 &=& \sum_l l^2 p_{k,l}(y) -m^2 = k (1-4y^2)/4.
\end{eqnarray}
\begin{figure}[t]
  \centerline{\epsfig{file=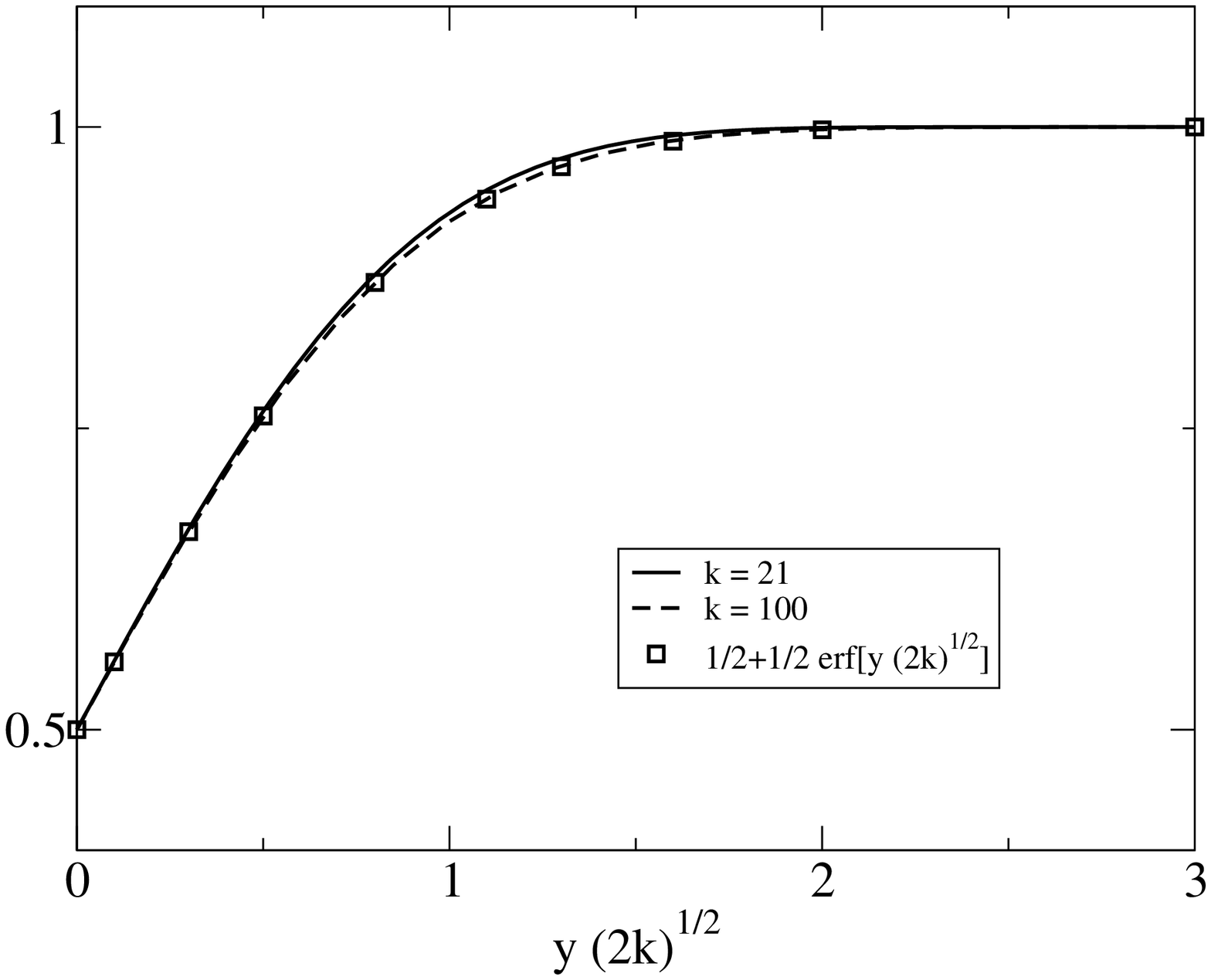, width=9cm}}
  \caption{Plot of $\Phi_k(1/2+y)$ as a function of $y \sqrt{2 k}$
    for different values of $k$ (lines)
    and the corresponding approximate form Eq.~(\ref{eq:20})
    (square symbols).} 
  \label{phi}
\end{figure}
If we insert this form into the definition of  $\Phi_k(1/2+y)$, and
substitute the sum over $l$ by an integral, we obtain
\begin{equation}
   \Phi_k(1/2+y) =  \frac{1}{2} \mathrm{erf}\left( 
       \frac{y \sqrt{2 k}}{(1-4 y^2)^{1/2}} \right)
     +
     \frac{1}{2} \mathrm{erf}\left( 
       \frac{\sqrt{2 k}(1/2-y)}{(1-4 y^2)^{1/2}} \right),
     \label{eq:19}
\end{equation}
where $\mathrm{erf}(x)$ is the error function \cite{abramovitz}. For
constant $y \sqrt{k}$ and large $k$, the argument of the second term in
Eq.~(\ref{eq:19}) becomes large, and we can approximate this term by unity.
In the same limit, the $y$ contribution in the
denominator of the argument in the first term of this equation is negligible,
so that
\begin{equation}
   \Phi_k(1/2+y) = \frac{1}{2} + \frac{1}{2} \mathrm{erf}\left(y \sqrt{2
       k} \right). 
   \label{eq:20}
\end{equation}
In Fig.~\ref{phi} we check the validity of this analytical expression
by plotting it agains the result of the numerical evaluation of
Eq.~(\ref{eq:21}) as a function of the rescaled variable $y \sqrt{2
  k}$.

\begin{table}[b]
  \caption{Functional forms of $\psi(y)-1/2$ as a function of $y$ for
  the different ranges of $\gamma$.}
  \label{Table1}
  \begin{center}
    \begin{tabular}{c|c|c}
      &$2 < \gamma < 5/2$ & $\gamma>5/2$ \\\hline
      $y \ll y_c$ &$c_2(\gamma) k_0^{\gamma-2} k_c^{5/2-\gamma} y$
      &$c_1(\gamma) k_0^{1/2} y$ \\
      $y_c \ll y \ll y_{NL}$ & $c_3(\gamma) k_0^{\gamma-2} y^{2(\gamma-2)}$
      &$c_1(\gamma) k_0^{1/2} y$ \\
      $y \gg y_{NL}$ &$1/2$ & $1/2$
    \end{tabular}
  \end{center}  
\end{table}

Inserting the functional form Eq.~(\ref{eq:20}) into the definition of
$\psi(y)$, we obtain, within the continuous degree approximation, and
neglecting terms of order $(k_c/k_0)^{1-\gamma}$,
\begin{equation}
  \psi(y) \simeq  \frac{1}{2} + \frac{(\gamma-2) k_0^{\gamma-2}}{2} \int_{k_0}^{k_c}
  k^{-\gamma+1}  \mathrm{erf}\left(y \sqrt{2 k} \right) dk.
  \label{eq:11}
\end{equation}
The behavior of this function has different asymptotic regimes,
depending on the value of $y$. In the first place, if $\sqrt{2 k_c} y$
is small, then the argument in the error function is small for all
values of $k$, and we can substitute it by its linear approximation,
$\mathrm{erf}(x) \simeq 2x/ \sqrt{\pi}$. Then, for $y< y_c = k_c^{-1/2}$, we
have
\begin{eqnarray}
  \psi(y) &\simeq&  \frac{1}{2} + \frac{\sqrt{2}(\gamma-2) k_0^{\gamma-2}}{\sqrt{\pi}} y
  \int_{k_0}^{k_c}   k^{-\gamma+3/2}dk\\ 
  &=& \frac{1}{2} + \frac{\sqrt{2}(\gamma-2) k_0^{\gamma-2}}{\sqrt{\pi}} y 
  \frac{k_c^{5/2-\gamma} - k_0^{5/2-\gamma}}{5/2-\gamma}.
\end{eqnarray}
As we can observe, the behavior of $\psi(y)$ for $y<y_c$ depends now also
on the value of the degree exponent $\gamma$.  For $\gamma>5/2$, the
contribution of $k_c$ vanishes in the limit of large $k_c$, and the
function $\psi(y)$ takes the functional form $\psi(y) \sim 1/2 + c_1(\gamma)
k_0^{1/2} y$. For $\gamma<5/2$ on the other hand, the most relevant
contribution comes from the degree cut-off, and we have $\psi(y) \sim 1/2 +
c_2(\gamma) k_0^{\gamma-2} k_c^{5/2-\gamma} y$. Here $c_1(\gamma)$ and $c_2(\gamma)$ are
constants that depend only on the degree exponent $\gamma$.

When $y>y_c$, we must take into account the full functional form of
the error function in order to evaluate Eq.~(\ref{eq:11}). In this case
we have that the argument of the error function is large at the upper
limit of the integral. Given that the error function saturates to one
for large argument, we can simplify the integral extending its upper
limit to infinity. Therefore, we have that, in this range of $y$
values,
\begin{equation}
  \psi(y) \simeq  \frac{1}{2} + \frac{(\gamma-2) k_0^{\gamma-2}}{2} \int_{k_0}^{\infty}
  k^{-\gamma+1}  \mathrm{erf}\left(y \sqrt{2 k} \right) dk.
\label{eq:12}
\end{equation}
This last integral can be expressed analytically  in
terms of the incomplete Gamma function~\cite{abramovitz}.  Performing
an expansion of the resulting expression for $y \ll y_{NL} =
k_0^{-1/2}$, we obtain again two limiting regimes, depending on the
degree exponent.  For $\gamma > 5/2$, the leading term is again $\psi(y) \sim 1/2
+ c_1(\gamma) k_0^{1/2} y$, that is, we obtain the same linear behavior as
in the case $y< y_c$.  For $2<\gamma<5/2$ instead, the integration over $k$
brings a singular behavior, so that we obtain $\psi(y) \sim 1/2 + c_3(\gamma)
k_0^{\gamma-2} y^{2(\gamma-2)}$.  Here $c_3(\gamma)$ is again a constant that depends
only on the degree exponent $\gamma$.  Finally, for $y \gg y_{NL}$ the
argument in the error function in Eq.~(\ref{eq:12}) is large for all
$k$, and we can approximate $\mathrm{erf}(x) \simeq 1$. In this case,
$\psi(y)$ can be taken to be equal to $1$ for any value of $\gamma$.  The
behavior of the function $\psi(y)$ is summarized in Table~\ref{Table1}.

\providecommand{\newblock}{}


\end{document}